\documentclass[manuscript]{acmart}
\newcommand{\multicomment}[1]{}

\AtBeginDocument{%
  }

\setcopyright{acmlicensed}
\copyrightyear{2026}
\acmYear{2026}
\acmDOI{XXXXXXX.XXXXXXX}

\acmBooktitle{ACM Transactions on Software Engineering and Methodology}
\acmISBN{978-1-4503-XXXX-X/18/06}

\usepackage{amsmath,amsfonts}
\usepackage{graphicx}
\usepackage{textcomp}
\usepackage{caption}
\usepackage{blindtext}
\usepackage{multicol}
\usepackage{xcolor}
\usepackage[table]{xcolor} %
\usepackage{tikz}
\usepackage{listings}
\usepackage{enumitem}
\usepackage{hyperref}
\usepackage{amsfonts}
\usepackage{wrapfig}
\usepackage{tcolorbox}
\usepackage{lipsum}
\usepackage{xspace} %
\usepackage{subcaption} 
\usepackage{adjustbox}
\usepackage{fancybox}

\usepackage[noend]{algpseudocode}
\usepackage[ampersand]{easylist}

\usepackage{colortbl}
\usepackage[normalem]{ulem}  %
\useunder{\uline}{\ul}{}  %
\usepackage{booktabs} %
\usepackage{multirow} %
\usepackage{tabularx}      %

\newcommand{\myp}[1]{\noindent\emph{\textbf{#1}}}

\newcommand{\ie}{\textit{i.e.,}\xspace}
\newcommand{\eg}{\textit{e.g.,}\xspace}

\newcommand{\etal}{et al.\xspace}

\newcommand{\randomCut}{\textit{RandomCut}\xspace}

\newcommand{\dvocab}{\texttt{d\_vocab\_size}\xspace}

\newcommand{\ASTerrors}{\texttt{w\_n\_ast\_errors}\xspace}
\newcommand{\ASTlevels}{\texttt{w\_n\_ast\_levels}\xspace}
\newcommand{\ASTnodes}{\texttt{w\_n\_ast\_nodes}\xspace}
\newcommand{\whitespaces}{\texttt{w\_n\_whitespaces}\xspace}
\newcommand{\complexity}{\texttt{w\_complexity}\xspace}
\newcommand{\nloc}{\texttt{w\_nloc}\xspace}
\newcommand{\tokenCount}{\texttt{w\_token\_count}\xspace}
\newcommand{\identifiers}{\texttt{w\_n\_identifiers}\xspace}
\newcommand{\commitID}{\texttt{commit\_id}\xspace}
\newcommand{\funName}{\texttt{fun\_name}\xspace}
\newcommand{\filename}{\texttt{file\_name}\xspace}
\newcommand{\code}{\texttt{code}\xspace}
\newcommand{\dwhitespaces}{\texttt{w\_d\_whitespaces}\xspace}
\newcommand{\commitMessage}{\texttt{commit\_message}\xspace}
\newcommand{\docstring}{\texttt{docstring}\xspace}

\newcommand{\iwhitespaces}{\texttt{i\_n\_whitespaces}\xspace}
\newcommand{\inwords}{\texttt{i\_n\_words}\xspace}
\newcommand{\ivocab}{\texttt{i\_vocab\_size}\xspace}
\newcommand{\ewhitespaces}{\texttt{e\_n\_whitespaces}\xspace}
\newcommand{\enwords}{\texttt{e\_n\_words}\xspace}
\newcommand{\wnwords}{\texttt{w\_n\_words}\xspace}
\newcommand{\evocab}{\texttt{e\_vocab\_size}\xspace}
\newcommand{\wvocab}{\texttt{w\_vocab\_size}\xspace}
\newtcolorbox{boxK}{
    fontupper = \small,
    sharpish corners, %
    boxrule = 0pt,
    toprule = 0pt, %
}

\newcommand{\secref}[1]{Sec.~\ref{#1}\xspace}
\newcommand{\figref}[1]{Fig.~\ref{#1}\xspace}
\newcommand{\tabref}[1]{Tab.~\ref{#1}\xspace}

\newcommand{\docode}{\textit{do$_{code}$}\xspace}
\newcommand{\causalse}{\textit{CausalSE}\xspace}
\newcommand{\ese}{\textit{ESE}\xspace}

\newcommand{\llms}{\textit{LLMs}\xspace}
\newcommand{\llm}{\textit{LLM}\xspace}

\newcommand{\gpt}{\textit{GPT-3}\xspace}
\newcommand{\scm}{SCM\xspace}
\newcommand{\scms}{SCMs\xspace}
\newcommand{\docalculus}{\textit{do-calculus}\xspace}

\newcommand{\da}{\textit{DAG}\xspace}
\newcommand{\dags}{\textit{DAGs}\xspace}

\begin{document}

\title{Rethinking Software Empirical Studies with Structural Causal Models
}

\author{Daniel Rodriguez-Cardenas}
\authornote{Authors contributed equally.}
\email{dhrodriguezcar@wm.edu}
\orcid{1234-5678-9012}
\author{Aya Garryyeva}
\authornotemark[1]
\email{lgarryyeva@wm.edu}
\author{David N. Palacio}
\email{davidnad@microsoft.com}
\authornotemark[1]
\author{Antonio Mastropaolo}
\email{amastropaolo@wm.edu}
\author{Denys Poshyvanyk}
\email{dposhyvanyk@wm.edu}
\affiliation{%
  \institution{William \& Mary}
  \city{Williamsburg}
  \state{VA}
  \country{USA}
}

\renewcommand{\shortauthors}{Rodriguez-Cardenas, Garryyeva et al.}

\begin{abstract}

 Causal Inference offers a fundamental approach for advancing empirical software engineering (\ese) beyond traditional statistical association, enabling researchers to rigorously identify and quantify causal relationships in software experiments. This paper introduces \causalse, a framework that operationalizes Judea Pearl’s causal inference paradigm in ESE context. The paper focuses on Structural Causal Models (SCMs) to address the limitations of classical statistical methods in mitigating \textit{confounding bias}. Through a case study using the Galeras dataset and propensity score matching, we demonstrate how \causalse disentangles the effect of prompt engineering strategies on code generation outcomes in a popular \llm (\ie \gpt). The results reveal that while associational analyses can suggest improvements in certain interventions (\eg more complex prompts), causal analysis often does not find a significant treatment effect, highlighting the risk of false positives when confounding is not addressed. By providing a tutorial-based methodology and a real-world case study, this work equips software researchers with practical tools to design, analyze, and interpret software experiments with methodological rigor, ultimately enabling more informed and actionable conclusions in both research and practice.
\end{abstract}

\ccsdesc[500]{Mathematics of computing~Hypothesis testing and confidence interval computation}
\ccsdesc[500]{Mathematics of computing~Probabilistic representations}

\keywords{Causal Graphs, Empirical SE, Spurious Correlation}

\received{20 February 2007}
\received[revised]{12 March 2009}
\received[accepted]{5 June 2009}

\maketitle

\section{Introduction}
\label{sec:intro}

Scientific \textit{explanation} and \textit{causality} are deeply intertwined: for most philosophers of science, explaining a phenomenon is to identify what caused it. The debate traces back to Aristotle (322 BCE), who organized scientific inquiry around \texttt{why-type} questions, and to Hume (1775), who countered that we never observe causation directly; only the habitual conjunction of events. To this day, no single definition of causality has received universal acceptance, either philosophically or scientifically.

Contemporary authors argue for Humean reduction. Cartwright \cite{cartwright1989capacities,cartwright1999dappled,cartwright2007hunting} and Bunge \cite{bunge1959causality,bunge2003emergence,bunge2011philosophy} argue that causation must be grounded in \textit{mechanisms} rather than statistical correlations\footnote{Bunge defines a \textit{``mechanism''} as a process in a concrete system whereby one state or event brings about another under laws of nature \cite{bunge2011philosophy}.}. The two diverge on the details: Cartwright admits \textit{context-dependent} causation in place of strict deterministic laws, while Bunge retains determinism but embeds it in complex, multilayered system interactions\footnote{For Bunge, the world is organized as systems nested within systems; \textit{``multilayered interactions''} mean that events at one level \textit{affect} and \textit{are affected by} events at others \cite{bunge2011philosophy,bunge2003emergence,bunge1959causality}. A causal chain may therefore span layers, \eg gene mutation $\rightarrow$ brain chemistry $\rightarrow$ personality $\rightarrow$ economic decision.}. Judea Pearl operationalized this mechanistic view under the banner of \textbf{causal inference}, turning what had been a philosophical commitment into a computational and mathematical framework to pose and answer causal queries
\cite{Pearl2018Causality,Pearl2009Causality,pearl2009overview}. Cartwright and Bunge have, in turn, lent both support to causal inference's logic and critique of its practical reach.

\begin{wrapfigure}{r}{0.3\linewidth}
 \centering
 \includegraphics[width=\linewidth]{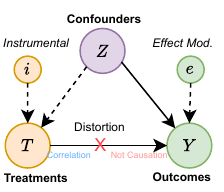}
 \caption{SCM variables.}
 \label{fig:causalgraph}
\end{wrapfigure}

At the heart of Pearl's framework lies \docalculus, a symbolic engine that translates the effect of an action into an expression over observed data \cite{Pearl2009Causality}. This machinery is what gives teeth to the aphorism \textit{``correlation is not causation''}, formalizing the gap between observed associations and true causal effects. That gap has two principal sources: \textbf{selection bias}, which arises when data are conditioned on common outcomes, and \textbf{confounding bias}, which arises from common causes; variables $Z$ that influence both treatments $T$ and outcomes $Y$ (\figref{fig:causalgraph}). In \docalculus, the \textit{do-operator} writes interventions as $p(y\mid do(t))$ and contrasts them with the ordinary conditional $p(y\mid t)$. Then confounding is the formal statement that the two disagree: $p(y\mid t)\neq p(y\mid do(t))$.

Pearl arranges causal queries on a hierarchy known as the \textit{ladder of causation}, with three rungs of increasing expressive power: \textit{L$_1$-association} (``seeing''), \textit{L$_2$-intervention} (``doing''), and \textit{L$_3$-counterfactuals} (``imagining'')\cite{Pearl2018Causality}. Each rung answers questions the rungs below cannot (\tabref{tab:causal-hierarchy}). The \textit{association} rung captures passive statistical relationships $p(y\mid t)$; say, the correlation between code complexity and defect rates~\cite{kazman_causal_2017}, and is the home of classical \textit{correlation} and \textit{regression}. However, it cannot answer a \textit{``doing''} question such as \textit{``What if enforcing code reviews reduced bugs?''}, which requires us to actively \textit{change} the system. Such interventional queries reside on the second rung and are addressed through A/B tests, randomized control trials, instrumental variables, or simulation on \textit{causal graphs} via $p(y\mid do(t))$ \cite{molak_causal_2023,weinberg_causality_2024,pearl2009overview}; in our running example, $p(\textit{Bugs}\mid do(\textit{Code Reviews}))$. The third rung, \textit{counterfactuals}, supports retrospective reasoning of the form $p(y_t\mid t',y')$: the probability we would have seen outcome $Y=y$ under treatment $t$, given that we in fact observed $t'$ and $y'$. Counterfactuals subsume the lower rungs by combining observation with hypothetical intervention, answering \textit{``What would have happened?''} queries such as \textit{``Would the number of bugs be lower if we had used a different testing framework?''}~\cite{Pearl2016Causality}.

\begin{table*}[h]
\centering
\scalebox{0.95}{
\begin{tabular}{llll}
\toprule
\textbf{Level} &
  \textbf{Enables} &
  \textbf{Typical Questions} &
  \textbf{Research Question Examples} \\ \hline
\begin{tabular}[c]{@{}l@{}}\textbf{L$_3$-Counterfactuals }\\  
$p(y_t|t',y')$
\end{tabular} &
  \begin{tabular}[c]{@{}l@{}}Imagining, \\ Retrospection\end{tabular} &
  \begin{tabular}[c]{@{}l@{}}Why? \\ Was it $t$ that caused $y$? \\ What if I had acted differently?\end{tabular} &
  \begin{tabular}[c]{@{}l@{}}Would test growth continue \\ under alternative quality \\ gates?\end{tabular} \\
  
  \hline
\textbf{\begin{tabular}[c]{@{}l@{}}L$_2$-Intervention \\  %
$p(y|do(t))$
\end{tabular}} &
  \begin{tabular}[c]{@{}l@{}}Doing, \\ Intervening\end{tabular} &
  \begin{tabular}[c]{@{}l@{}}What if? What would happen \\ if we change $T=t$?\end{tabular} &
  \begin{tabular}[c]{@{}l@{}}What test automation policies \\ yield maximum ROI?\end{tabular} \\ 
  \hline
    \textbf{\begin{tabular}[c]{@{}l@{}}L$_1$-Association \\  %
    $p(y|t)$
    \end{tabular}} &
      \begin{tabular}[c]{@{}l@{}}Seeing\\ Observing\end{tabular} &
      \begin{tabular}[c]{@{}l@{}}What is? \\ How are variables related?\\ What patterns exist in our data?\end{tabular} &
      \begin{tabular}[c]{@{}l@{}}How does testing evolve \\ after continuous integration (CI) \\ adoption?\end{tabular} \\ 
  \bottomrule
\end{tabular}%
}
\caption{The Ladder of Causation with SE Examples.}
\label{tab:causal-hierarchy}
\end{table*}

The formal vehicle for climbing this ladder is the \textbf{Structural Causal Model}~(\scm), an extended form of causal graph that has become the backbone of cause-and-effect reasoning in modern AI. An \scm has three components: \textit{graphical models}, Directed Acyclic Graphs (\dags) that encode assumptions about the environment; \textit{counterfactual and interventional logic}, which supplies the semantics for articulating causal questions; and \textit{structural equations}, which tie the graph to the logic and determine when a counterfactual probability is estimable from experimental or observational data
\cite{pearl-theoretical-2018}.

In Empirical Software Engineering (\ese), \textbf{Structural Causal Model}~(\scm) let researchers move past surface-level correlations and reason about the mechanisms that drive software processes. They provide the formal language for a methodological question that recurs in software research: \textit{why should causal inference be integrated into empirical software experiments?} The answer, we argue, is that \scms enable a \textit{progression that classical statistics cannot support on its own}; from descriptive analysis (\textit{what is}) to interventional inquiry (\textit{what if}), and ultimately to retrospective reasoning (\textit{what would}). Each step deepens the analytical leverage available to software experts and turns empirical data into more reliable decision support. This paper focuses on the \textit{interventional} rung. We formalize the use of \scms to simulate interventions in software engineering, providing a rigorous methodological path to reasoning about causality in complex \ese settings.

The need for this shift is concrete. Classical statistical methods are \textit{causally opaque}: they detect associations but cannot, by design, separate genuine causal effects from spurious\footnote{\ie fictitious, illusory, apparent \cite{molak_causal_2023}.} ones without explicit causal assumptions \cite{pearl2009overview}. However, many of the questions that matter in software engineering, identifying the root causes of system failures \cite{kucuk_improving_2021,pham_root_2024,baah_causal_2010}, understanding how programming language choice affects competition performance and code quality\cite{furia_towards_2024,furia_mitigating_2025}, or testing whether developer headcount drives bug counts \cite{kazman_causal_2017}, are precisely of this kind. Relying on associational patterns alone risks misleading conclusions and, worse, software practices built on them. We therefore advocate moving from \textit{descriptive statistical empiricism} to \textit{causal-centric} software experimentation, which entails posing what-if questions and simulating interventions on top of causal explanations. The central question we address on this work is: \textit{how can software researchers incorporate causal inference into their empirical evaluations?}

Our hypothesis is that explicitly addressing confounding bias through causal inference strengthens the methodological rigor of empirical software experiments. Concretely, we state that \scms and \docalculus provide the mathematical foundation to (1)~\textbf{identify} confounding paths in the software data-generating process, (2)~\textbf{block} the spurious associations they induce through adjustment, and (3)~\textbf{quantify} interventional software-based effects that reflect true causal relationships rather than confounding artifacts.

To this end, we introduce \causalse, an \ese-specific framework for modeling, estimating, and validating causal relationships. Built on Pearl's causality, \causalse provides an extensible foundation that can target a range of causal properties; here we instantiate it to mitigate the impact of \textit{spurious correlations} and produce reliable causal explanations of complex software phenomena. We demonstrate \causalse on \textit{Galeras}, a dataset of code predictions \cite{daniel_icsme23}, showing how it explains the effect of prompt engineering (treatments) on generated code (outcomes). The instantiation is a rigorous adaptation of \docode~\cite{nader_palacio_toward_2024}, broadened from a single post-hoc interpretability setting to general \ese experimentation. Concretely, \causalse \textbf{extends} the scope of \docode from LLM explanation to general \ese experimentation and \textbf{adds} three SE-specific stages. A preliminary stage (Stage$_0$) elicits covariates and encodes domain knowledge before any modeling begins. Stage$_1$ tailors the modeling pipeline to code-structured data, with causal discovery, data transformation, and formal graph vetting. Stage$_5$ produces causal explanations by combining associational and interventional comparisons with robustness tests (refuters).

To the best of our knowledge, this paper responds to three idiosyncrasies of software data that generic causal frameworks do not handle: \textbf{observational confounding} in heterogeneous SE datasets, where hidden factors entangle actions and outcomes; \textbf{treatment design} on software-centric artifacts such as prompts, code comments, or vulnerabilities; and \textbf{multimetric outcomes} that span LLM-oriented metrics (\eg CodeBLEU, Levenshtein distance, perplexity) and software-specific ones (\eg bug count, code coverage, maintainability index).

We evaluate the framework with a descriptive case study on prompt engineering for code generation (\secref{sec:case}), which yields two methodological findings. First, \causalse isolates treatment effects in software data: it separates the influence of prompt context on \llm code-generation performance from the noise of confounding snippet attributes (\eg complexity, code length, and identifier count) that shape both the prompt strategy and the generated output. Formalizing the causal question \textit{``to what extent does prompt context influence the code-completion performance of an \llm, holding snippet-level attributes fixed?''}, \causalse recovers a treatment-effect estimate that associational analysis alone could not produce. Second, although associational analysis suggests that larger prompts improve code-generation performance, the causal analysis finds no statistically significant treatment effect: the observed correlations are driven by the above confounders, not by the prompt modifications themselves. This shows that \causalse can \textit{prevent the adoption of ineffective software practices by distinguishing genuine causal signals from confounding-induced artifacts}.

In light of these findings, the contributions of this work are as follows:
\begin{itemize}
    \item \textbf{SE-specific causal stages [\secref{sec:approach}]}. We extend standard causal estimators with three domain-specific stages: (i)~a \textit{Preliminary} stage that turns software expertise into candidate covariates and initial \scm{} templates; (ii)~a \textit{Modeling} stage with graph vetting and code-aware data processing that transforms raw code and documentation into valid treatments and outcomes; and (iii)~an \textit{Explanation} stage that contrasts associational and interventional results and validates effects with refuters. As illustrated in \figref{fig:causalSEpipeline}, these stages re-purpose the LLM-specific pipeline of $do\_{code}$ to the broader needs of empirical software research.
    \item \textbf{Case-study demonstration [\secref{sec:case}]}. We provide a detailed causal evaluation on a real-world software dataset that answers \textit{``to what extent does prompt context influence the code-completion performance of an \llm?''} while holding snippet-level confounders (complexity, code length, identifier count) fixed. The study shows how \causalse isolates treatment effects from the noise of the data set, recovers estimates that associational analysis cannot, and reveals actionable insights; such as drooping prompt-engineering correlations that vanish once confounding is controlled~\cite{RodriguezCardenas2026Causal4SE}.
    \item \textbf{Open framework and tutorial}. We release a comprehensive tutorial that integrates causal analysis into \ese experiments, with practical guidance and an open-source implementation to lower the entry barrier for the software engineering community.
\end{itemize}

By coupling a structured tutorial with open-source tooling, we aim to facilitate the transition from traditional association studies to rigorous structural causal analysis in \ese. Formalizing the steps of causal analysis lets researchers reason systematically about hypothetical interventions and alternative scenarios \cite{weinberg_causality_2024}, and evaluating multiple such scenarios sidesteps a structural limitation of empirical software studies: the impossibility of observing the same object of study (\eg a code generator, developers, or agents) under different treatments (\eg prompt engineering techniques, software methodologies, or policies) within a single experiment~\cite{siebert_applications_2023}.

\section{Causal Inference Background} %
\label{sec:background}

The foundational concepts introduced in this section lay the groundwork for examining the core components of causal inference, which we explore through three interconnected elements: formal definitions of causality, practical techniques for identification and estimation, and the confounding bias problem.

\textbf{Causal Concepts and Formal Definitions.} Causal analysis encompasses two complementary processes: causal discovery and causal inference. \textit{Causal discovery} involves learning the underlying causal structure from the data, essentially mapping the causal relationships between variables to determine which variables act as causes and which act as effects~\cite{molak_causal_2023, furia_towards_2024}. \textit{Causal inference}, in turn, focuses on quantifying causal effects once the causal structure is known or assumed~\cite{siebert_applications_2023, weinberg_causality_2024}. Although various causal modeling frameworks exist, such as Rubin's potential outcomes, this work relies on the Pearlian \textit{GCM} framework. As Siebert notes, ``Statistical causal inference focuses on estimating the actual causal effects of an action (a treatment $T$) on a given observed system (an outcome $Y$) from data''~\cite{siebert_applications_2023}. A \textit{treatment} refers to ``the action that applies (exposes or subjects) to a unit''~\cite{yao_survey_2021}. Treatments are deliberately assigned in experimental settings, while in observational studies they occur naturally~\cite{rosenbaum_choice_1999, hernan_causal_2020}.

\textit{Causal diagrams} make observable assumptions about the relationships between variables, enabling rigorous critique and refinement~\cite{Pearl2016Causality}. Pearl presents \dags as a unified language and toolkit to assess when causal conclusions can be drawn from observational data~\cite{pearl-seven-2019}. In a \da, the variables of interest are represented as nodes and their causal dependencies as directed edges, allowing practitioners to visualize and communicate causal assumptions explicitly. \dags specify a set of exogenous variables (external noise terms) and endogenous variables whose values are determined by structural functions of their direct causes~\cite{Pearl2016Causality}. Because every directed edge encodes an asymmetric functional relationship, these arrows both confirm an association and establish which variable is the cause~\cite{molak_causal_2023}. A directed edge $T \to Y$ indicates that $T$ appears in the structural function that assigns the value to $Y$~\cite{Pearl2016Causality}; therefore, variable $T$ \textit{directly causes} variable~$Y$.

\textit{Confounding} arises when a variable simultaneously influences both the treatment and the outcome, thereby biasing the estimation of causal effects~\cite{molak_causal_2023}. Pearl characterizes confounding as a fundamental obstacle that conflates \textit{seeing} (observing associations) with \textit{doing} (intervening on causes)~\cite{Pearl2018Causality}. Confounding bias thus occurs when such variables distort the treatment-outcome relationship. \scm address this problem by blocking \textit{backdoor} paths through covariate adjustment, governed by the criterion of \textit{d-separation}~\cite{Pearl2018Causality}. The backdoor criterion identifies a set of variables $Z$ that, when conditioned upon, block all spurious non-causal pathways between $T$ and $Y$. If $T$ and $Y$ are d-separated given $Z$ in a \da, then $T$ is conditionally independent of $Y$ given $Z$.

Building on these structural foundations, \textit{interventions} in Pearl's framework are formalized through the $do$-operator, denoted $do(T=t)$~\cite{pearl-theoretical-2018}. Whereas adjustment sets block spurious paths by conditioning on covariates, the $do$-operator simulates the act of setting variable $T$ to a specific value $t$, severing it from the influence of its natural causes. This distinction elevates causal analysis from passive observation to active experimental reasoning, enabling investigators to estimate interventional effects from observational data by reasoning about hypothetical manipulations or changes~\cite{weinberg_causality_2024}.

\textbf{Practical Techniques for Identification and Estimation.} With the structural foundations of the causal model established, we now turn to the methodologies required for quantifying causal effect -specifically, propensity score approaches and their application in SE contexts. The \textit{estimation} of causal effects involves three distinct components: the \textit{estimand}, which defines the target causal quantity (\eg ATE, CATE); the \textit{estimator}, which specifies the method applied (\eg propensity score matching); and the \textit{estimate}, which is the numerical result obtained by applying the estimator to the data~\cite{molak_causal_2023}. More succinctly, the estimand captures the \textit{what}, the estimator the \textit{how}, and the estimate the \textit{how much}~\cite{molak_causal_2023}. Pearl conceptualizes this estimation pipeline within the \scm framework as an \textit{inference engine}~\cite{pearl-seven-2019, pearl-theoretical-2018}.

The \textit{estimand} is a mathematical expression derived from the model's assumptions and the target causal query. Causal effects are quantified through five primary estimands: the average treatment effect (\textbf{ATE}), the average treatment effect on the treated (\textbf{ATT}), the average treatment effect on the control (\textbf{ATC}), the conditional average treatment effect (\textbf{CATE}), and the individual treatment effect (\textbf{ITE}). ATE measures the expected difference in outcomes between treatment and control throughout the population~\cite{yao_survey_2021}. ATT quantifies the effect among units that actually received treatment, while ATC captures the corresponding effect for untreated units~\cite{molak_causal_2023}. CATE extends ATE to the subgroup level, allowing the estimation of heterogeneous effects when the impact of treatment varies between subpopulations~\cite{yao_survey_2021}. At its finest granularity, ITE represents the inherently unobservable difference in potential outcomes for a single unit. Although ATE, ATT, and CATE are empirically estimable, ITE remains a latent quantity; nevertheless, it serves as the theoretical foundation for personalized interventions~\cite{yao_survey_2021}.

Once causal assumptions are formalized, \textit{identification methods}, also known as estimators, determine whether and how the target causal effect can be recovered from observational data~\cite{siebert_applications_2023}. Within the \scm framework, identification amounts to verifying that the estimand is derivable from the model's structural assumptions and the observed data distribution. Pearl's $do$-calculus provides the formal machinery for this purpose, systematically identifying and eliminating spurious associations introduced by confounding~\cite{siebert_applications_2023}. In practice, three propensity-score-based estimators are commonly employed to quantify causal effects from non-experimental data. \textit{Propensity Score Weighting (PSW)} constructs a synthetic pseudo-population in which the covariate distributions are balanced between treated and control groups, thus mimicking randomization and producing unbiased effect estimates provided that all confounders are measured~\cite{molak_causal_2023}. \textit{Propensity Score Stratification (PSS)} partitions the sample into strata based on estimated propensity scores; the treatment effect is estimated within each stratum and then aggregated between strata to produce an overall estimate. \textit{Propensity Score Matching (PSM)} pairs each treated unit with one or more control units that share similar propensity scores, constructing comparable groups that reduce confounding and enable a more accurate estimation of treatment effects from observational data.

The treatment effect estimates produced by propensity score methods indicate the direction and magnitude of the causal relationship between treatment and outcome. A \textbf{positive effect} indicates that the treatment is associated with an increase in the outcome, a \textbf{null effect} indicates that there is no meaningful causal association, and a \textbf{negative effect} indicates that the treatment decreases the outcomes. The choice among these estimators depends on the causal structure encoded in the graph and the strength of the assumptions warranted by domain knowledge. Together, these practical techniques are embedded within the causal inference pipeline, enabling SE practitioners to move from theoretical assumptions to quantifiable treatment effects while accounting for confounding bias.

\textbf{The Confounding Bias Problem in Empirical Software Engineering.} Confounding arises when common causes simultaneously influence both the treatment (\eg a development practice) and the outcome (\eg software quality), producing spurious associations that masquerade as causal effects. In \ese, where randomized controlled trials are often impractical or ethically infeasible, researchers must draw causal conclusions from observational data sources such as GitHub repositories, industrial codebases, and bug-tracking systems. In these non-experimental settings, confounding is an inherent characteristic of the data rather than a rare anomaly. The absence of standardized causal frameworks to identify and mitigate confounders has led to three critical consequences: (1)~\textit{false positives}, in which observed correlations are misinterpreted as causal effects without adequate adjustment for confounders; (2)~\textit{inconsistent findings}, where studies of the same phenomenon yield contradictory results due to uncontrolled confounding bias; and (3)~\textit{an inability to answer interventional questions}, \eg ``What would happen if we adopted practice~$X$?'', which practitioners require for evidence-based decision-making.

The study by Neto~\etal~\cite{soares_continuous_2023} provides a compelling illustration of confounding bias in continuous integration (CI) research. Although prior observational studies had routinely attributed improvements in software quality directly to CI adoption, their causal analysis revealed project age as a critical confounder mediating this relationship. Under a na\"ive correlation-based approach, one might erroneously conclude that adopting CI directly reduces defects, on the basis of strong correlations and high association values. However, the causal analysis by Neto~\etal clarifies that project maturity drives both CI adoption and improved software quality: mature projects typically have more established development processes and more experienced contributors; factors that inherently elevate quality independent of CI~\cite{soares_continuous_2023}. From a practical standpoint, recognizing this confounding structure explains why some development teams fail to observe the anticipated quality improvements after adopting CI and underscores the necessity of causal reasoning when evaluating the efficacy of development practices.

Integrating causal inference into \ese addresses long-standing limitations of purely statistical approaches. By explicitly modeling causal mechanisms, researchers can systematically mitigate biases arising from confounding variables. In the pioneering study \textit{Applications of Statistical Causal Inference in Software Engineering}, Siebert~\cite{siebert_applications_2023} provides the first systematic mapping of causal inference methods to the software engineering domain. However, existing syntheses, including Siebert's mapping~\cite{siebert_applications_2023}, advocate for standardized and reproducible causal methodology in SE but stop short of providing a procedural pipeline for practitioners. Similarly, Ji~\etal~\cite{ji_benchmarking_2023} demonstrate the application of \scms to LLM-based code generation, a contribution that remains narrowly scoped to that task rather than addressing general \ese experimentation. \causalse fills this gap by operationalizing an end-to-end workflow that (a)~structures pre-modeling elicitation for software data, (b)~instantiates and validates \scms through code-aware processing, and (c)~assesses interventional conclusions via refutation tests and an explicit contrast between associational and interventional estimates.

By surveying emerging applications, primarily in software quality analysis and testing, Siebert highlights that the field remains fragmented, revealing clear opportunities to standardize causal methodologies and practices. Adopting causal inference techniques therefore equips software engineering researchers and practitioners not only with the ability to diagnose problems, but also to design targeted interventions. These interventions transform observational data into evidence-based guidelines that can significantly improve software development practices and processes.

To close this methodological gap, we introduce the \causalse framework, which operationalizes Pearl's causal inference paradigm specifically for \ese contexts. \causalse addresses confounding bias through three complementary mechanisms: (1)~explicit \scm construction that encodes confounding assumptions in \dags, enabling systematic identification of biasing pathways; (2)~formal identification strategies, including the backdoor criterion and instrumental variables, that determine when and how to adjust for confounders; and (3)~estimation and validation pipelines, comprising propensity score methods and refutation tests, that quantify causal effects while stress-testing their robustness to hidden confounding. The remainder of this paper demonstrates how \causalse transforms confounding from an implicit threat into an explicit, manageable component of empirical software experiments.

\section{Causal Software Engineering: From \docode to \causalse}\label{sec:approach}

\begin{figure*}[ht]
  \begin{center} 
  \includegraphics[width=0.95\textwidth]{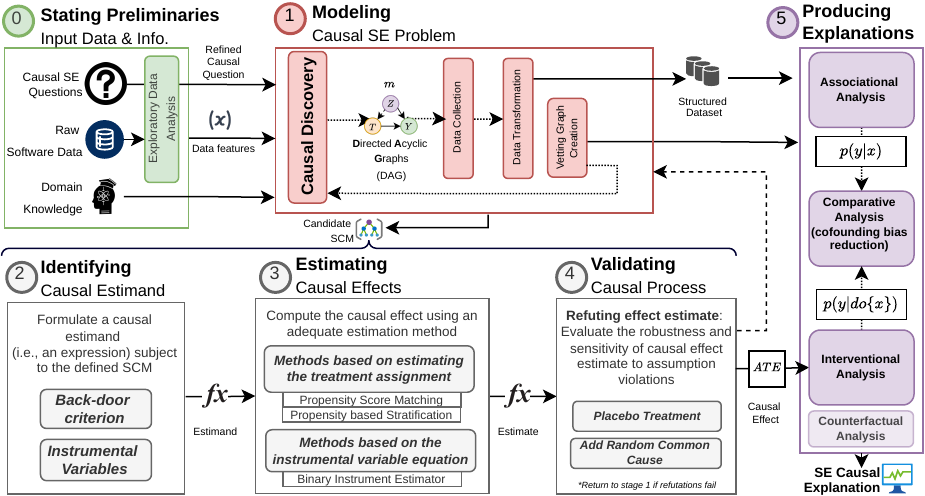}
  \end{center} 
  \caption{\causalse Pipeline. Boxes in color are SE-based adaptations for \docode (in gray). }
  \label{fig:causalSEpipeline} 
\end{figure*} 

This section presents the foundational components of \causalse, our proposed framework to enable causal analysis in empirical software engineering. We argue that strengthening the rigor of SE evaluations requires a set of causal tools designed to identify and mitigate \textit{confounding bias} among the covariates under study. We introduce these tools and highlight representative SE problems that stand to benefit from the proposed pipeline. A concrete, end-to-end demonstration of how each stage operates on real-world software data is deferred to \secref{sec:case}.

\subsection{The \docode Pipeline}\label{subsec:docode}

\docode is a \textbf{post-hoc interpretability method} designed to explain \llm code predictions through \textbf{causal queries}~\cite{nader_palacio_toward_2024}. By grounding its explanations in causal theory, the framework enables researchers and practitioners to identify model limitations, refine training datasets, and make more informed decisions when building automated developer tools. To mitigate the influence of spurious correlations, \docode anchors its analyzes in the structural and semantic properties of the underlying source code, ensuring that the behavior of the model is attributed to meaningful programming constructs rather than incidental statistical patterns.

The \docode pipeline comprises four major stages. In the first stage, \textbf{Structural Causal Graph Construction}, the relationships among variables are formalized as an \scm. This requires explicitly specifying the \textbf{interventions} ($T$), which typically represent input data characteristics or model configurations; for instance, a domain expert might investigate how the model's outputs are affected by specific code properties (\eg the presence of bugs, the density of inline comments, or the number of model layers). The stage also involves defining the \textbf{potential outcomes} ($Y$), which capture the aspects of \llm behavior under study, such as cross-entropy loss or next-token prediction performance. To complete the causal graph, researchers must identify the \textbf{confounders} ($Z$): common causes that simultaneously influence both the interventions and the outcomes. These confounders are typically derived from domain knowledge of \llm architectures and observable software properties, such as standard code quality metrics. All causal assumptions must be encoded explicitly in the graph.

In the second stage, \textbf{Causal Estimand Identification}, \docode translates the structural assumptions encoded in the graph into a formal estimation strategy. The framework employs graph-based criteria; including the backdoor criterion, frontdoor criterion, instrumental variables, and mediation analysis, to determine whether the causal effect is identifiable from the available data and which confounders must be adjusted for. Specifically, \docode applies adjustment formulas to select the precise set of variables ($Z$) that must be conditioned upon to isolate the true relationship between the intervention ($T$) and the outcome ($Y$). In this context, conditioning serves as a statistical control that blocks the influence of confounding pathways, ensuring that the estimated association is not distorted by common causes. This process transforms the theoretical relationships specified in the \scm into a target estimand computable from observational data.

The third stage, \textbf{Causal Effect Estimation}, applies statistical and machine learning methods to the observed software data to quantify the identified estimand. The choice of estimator depends on the nature of the variables within the \scm (\eg binary, discrete, or continuous). For binary SE interventions, such as comparing buggy versus fixed code, \docode employs propensity score matching. For discrete interventions (\eg hyperparameter variations) or continuous interventions (\eg fluctuating cyclomatic complexity scores), the framework uses linear regression to approximate the causal effect. The primary output of this stage is the \textit{Average Treatment Effect} (ATE), which quantifies the expected change in the outcome across the entire population when the intervention is modified, producing a single, interpretable measure of causal impact.

The final stage, \textbf{Causal Process Validation}, assesses the validity of the obtained causal estimates through two complementary mechanisms. First, \textit{refutation testing} subjects the findings to a series of sensitivity and robustness analyzes. This process does not presume the estimated effect is spurious; rather, it quantifies how much the estimate shifts when the underlying assumptions are deliberately challenged or violated; if the effect remains stable across these perturbations, it gains higher credibility. Second, \textit{graph vetting} employs correlational analyses on the \scm's variables to verify whether the assumed causal relationships are consistent with observed data patterns, thereby providing empirical support for the graph's structure. By examining the strength of associations between interventions, outcomes, and confounders, researchers can detect potential structural misspecifications in the causal model.

By integrating these four stages, \docode provides a rigorous end-to-end framework for causal analysis of large language models. This pipeline ensures that explanations are not only statistically sound but also grounded in the semantic realities of software development, enabling researchers to move beyond surface-level correlations toward a deeper, causally informed understanding of model behavior.

\subsection{\causalse Pipeline Adaptations}\label{subsec:adaptations}

\causalse extends the \docode framework~\cite{nader_palacio_toward_2024} for empirical software studies by introducing three domain-specific enhancements: (i)~a preliminary stage, Stage$_0$, that derives software-specific covariates and \scm hypotheses by combining domain knowledge with exploratory analysis; (ii)~an expanded Stage$_1$ incorporating causal discovery, data transformation, and graph-vetting mechanisms tailored to code and documentation artifacts; and (iii)~a reframed Stage$_5$ that produces SE-oriented causal explanations by contrasting associational results with interventional findings and validating effects through placebo and random-cause refuters. While the underlying identification, estimation, and validation algorithms follow the \docode pipeline described in \secref{subsec:docode}, all inputs, intermediate artifacts, and outputs are specialized for the SE domain. The colorful boxes in \figref{fig:causalSEpipeline} mark these additions.

By generalizing \docode to the analysis of software data, our aim is to help SE researchers and practitioners recognize the statistical limitations inherent in conventional software evaluations and advance toward more robust empirical analyses grounded in causal models. With this foundation, practitioners can make better-informed decisions about how to design and interpret empirical studies, guided by a holistic understanding of \textit{\textbf{why}} observed results occur rather than merely \textit{that} they occur.

We design the pipeline \causalse under the assumption that well-defined causal questions for \ese have been formulated by domain experts prior to analysis. Consequently, \causalse is structured to answer causal questions in the following form: \textit{why do specific software engineering interventions influence the observed system outcomes?} To address a given causal question, these \textit{interventions} represent software artifacts or practices identified by domain experts (\eg code bugginess, prompt engineering strategies). \causalse models the resulting causal relationships using the three fundamental graphical structures of causal analysis; chains, forks, and colliders~\cite{Pearl2016Causality}, and applies appropriate identification methods to isolate the true effects. These include the \textit{backdoor criterion} for blocking spurious paths between treatment and outcome, \textit{instrumental variables} for addressing confounding when common causes are unobserved, and the \textit{frontdoor criterion} for estimating causal effects through mediating variables that transmit the treatment's influence to the outcome.

In what follows, we detail the inputs, outputs, internal processes, and information flow of the three stages that \causalse adds to the \docode pipeline. These extensions are highlighted with colorful boxes within Stage$_0$, Stage$_1$, and Stage$_5$ of the pipeline diagram (\figref{fig:causalSEpipeline}). Stage$_2$ (identification), Stage$_3$ (estimation), and Stage$_4$ (validation) remain as originally described in \secref{subsec:docode}.

\textbf{Stage$_0$ [Adapted]: Stating Preliminaries.}
This stage integrates three types of input. First, researchers formulate causal research questions grounded in a specific software engineering problem; each question must be amenable to empirical investigation with available or collectible data. Second, raw software data relevant to the causal problem are gathered; either through new collection efforts or by reusing datasets from prior studies. From the research question and the collected data, practitioners identify potential \textit{covariates} that can influence causal relationships through \textit{Exploratory Data Analysis (EDA)}. When no initial dataset is available, domain expertise is used to qualitatively derive likely covariates. Third, domain knowledge is required to approximate the causal structure of the underlying \textit{data-generating process}, providing an initial structural model of the directional relationships among covariates. Stage$_0$ produces three outputs: a refined causal question, a data-feature profile (variables, distributional characteristics, and pairwise correlations) obtained through EDA, and a preliminary understanding of the causal structure that informs the subsequent modeling stage.

\textbf{Stage$_1$ [Adapted]: Modeling the Causal SE Problem.}
This stage receives two key inputs from Stage$_0$: (1)~the data features collected to identify potential covariates relevant to the causal question, and (2)~expert domain knowledge of the underlying causal structure. When empirical data are available, an EDA identifies an initial set of covariates that seeds the causal discovery process; when data are absent or scarce, experts propose observable covariates from theoretical or empirical understanding alone. The modeling stage proceeds through four substages: \textit{causal discovery}, \textit{additional data collection}, \textit{data transformation}, and \textit{graph vetting}.

\textbf{Causal discovery} is the process of algorithmically learning the structure of a causal graph from observational data. Rather than requiring experts to specify the complete \da a priori, causal discovery algorithms, such as score-based methods, test conditional independence relationships among the measured covariates to propose candidate \da structures. In the SE domain, causal discovery is particularly valuable because the space of plausible variable relationships is large and domain intuitions can be incomplete or misleading. The output of this substage is one or more candidate \dags that are consistent with the observed data distribution and serve as the starting hypothesis for the full \scm.

\textbf{Additional data collection} may be triggered when a candidate \da reveals unmeasured variables, such as previously unrecognized confounders or instrumental variables, that are necessary to achieve identifiability. At this point, practitioners return to the data sources (\eg version-control repositories, issue trackers) or conduct targeted \llm experiments to extract the missing features before proceeding.

\textbf{Data transformation} restructures the raw software artifacts into the typed causal variables required by the \scm: treatments ($T$), outcomes ($Y$), confounders ($Z$), instrumental variables, and effect modifiers. For code-based data, this includes computing syntactic metrics from abstract syntax trees, normalizing metrics across programming languages, and encoding categorical attributes as binary treatment indicators.

\textbf{\da vetting} is the complementary, expert-driven substage that guards against purely data-driven errors. Even when a candidate \da is statistically consistent with the observed data, it may encode implausible or impossible causal directions from a domain perspective. Vetting proceeds by: (i)~inspecting the sign and magnitude of pairwise Pearson and Spearman correlations among all \scm variables to verify that assumed directed edges are at least associationally supported; (ii)~checking whether edges violate known temporal or logical constraints in the software domain (\eg an outcome metric cannot cause a prior input or retroactively alter lines of code); and (iii)~iteratively pruning or redirecting edges until the graph is both statistically defensible and substantively coherent. Because even minor structural deviations can substantially alter the identified adjustment sets and downstream causal conclusions~\cite{hulseShakyStructuresWobbly2025}; this hybrid approach, data-driven discovery followed by expert-driven vetting, is essential for constructing a reliable \scm.

Stage$_1$ yields two primary outputs: (i)~a structured dataset containing the typed causal variables for every observation, and (ii)~a set of vetted \scm candidates, each represented as a \da, ready for identification in Stage$_2$.

\textbf{Stage$_5$ [Adapted]: Producing Causal Explanations.}
Causal explanation is the ultimate objective of empirical science. A rigorous evaluation must account for the underlying causal structure (\ie the \scm) that approximates an answer to an interventional query. Estimating a causal effect amounts to quantifying the interventional distribution $p(y \mid do(x))$; this quantity therefore embodies the explanation for the phenomenon under investigation.

Stage$_5$ operationalizes this objective through three coordinated analyses. First, an \textbf{associational analysis} summarizes the raw statistical relationships between treatments and outcomes; for example, by reporting bootstrapped means and rank correlations across prompt strategies and code-quality metrics. Second, an \textbf{interventional analysis} contrasts these associations with the ATE and CATE estimates produced in Stage$_3$, answering the ``what would happen if we changed the intervention?'' question while adjusting for confounders. Third, a \textbf{comparative analysis} highlights discrepancies between the two preceding analyses: when associational results suggest an improvement that the interventional estimates and refutation tests do not corroborate, \causalse issues a null-effect conclusion and recommends revisiting Stage$_0$ to revise the \scm or collect additional data. Together, these three components provide causal explanations grounded in interventional distributions and framed as an explicit associational-versus-interventional contrast, equipping practitioners with the evidence needed to distinguish genuine causal effects from confounding-driven artifacts.

\subsection{Potential Risks and Mitigation}\label{subsec:risks}

Applying causal inference to software engineering entails several methodological risks that extend beyond routine modeling errors. Chief among these is \textit{unobserved confounding}, in which hidden variables, such as developer experience or organizational culture, simultaneously influence both treatment and outcome; biasing effect estimates even when the observable portion of \scm is correctly specified. A second pervasive risk is \textit{selection bias}: reliance on observational data from platforms such as GitHub means that the analyzed projects are often disproportionately successful, well-maintained, or popular, and may not be representative of the broader software ecosystem. Employing \textit{causal discovery} strategies within SE could mitigate some of these risks by surfacing latent structure in the data; however, this remains a largely under-explored direction in the field.

Critically, causal discovery algorithms alone cannot guarantee a correct \da for the \scm. Even minor structural deviations, a single reversed or omitted edge, can substantially alter the identified adjustment sets and downstream causal conclusions~\cite{hulseShakyStructuresWobbly2025}. To address this fragility, \causalse assigns a central role to software engineering expertise throughout graph construction: domain experts define treatments (\eg prompt design strategies), outcomes (\eg code-completion quality), and confounders (\eg code complexity, code length, number of identifiers), grounding every structural assumption in substantive knowledge. The framework also incorporates explicit validation steps to assess candidate causal graphs and prune implausible edges. \causalse therefore operates as a hybrid of data-driven discovery and expert-driven vetting, ensuring that the resulting \scms are both statistically consistent with the observed data and substantively meaningful from a software engineering perspective.

\section{A \causalse Study: Evaluating Prompt Engineering for Code Generation}\label{sec:case}

This section illustrates the \causalse\ workflow through a causal analysis of how prompt engineering strategies impact the code generation quality of LLMs. Following the five-stage methodology presented in \secref{sec:approach}, we conduct a \textit{descriptive case study} that evaluates LLM experiments in software engineering. Specifically, we investigate whether source code attributes act as confounders -common causes that influence both prompt templates (treatments) and the quality of generated code (outcomes), as measured by Levenshtein distance.

Our case study starts by establishing the preliminaries, drawing on the authors' expertise in SE to formulate the causal question and select the relevant raw data. The subsequent stages of the \causalse pipeline are explained in the following subsections and involve modeling the causal problem, identifying the causal estimate, validating the causal process, and generating causal explanations based on \docode.

We use prompt engineering to demonstrate how \causalse (i) generates artifact-level treatments and outcomes, (ii) constructs and validates a structural causal model (\scm), (iii) estimates ATE and CATE using propensity-based estimators, (iv) refutes the estimated effects, and (v) contrasts the results with associational analyzes. This workflow is designed to be replicable across diverse software engineering contexts. In this section, we detail the inputs, procedures, and outputs of each stage and discuss the corresponding steps of \causalse.

\subsection{[Stage$_0$] \causalse Preliminaries}\label{sec:stage0}

\myp{Inputs.} The preliminaries require first the raw \ese data, second a causal SE question, and third the practitioner's domain knowledge (\figref{fig:causalSEpipeline}). Our \textit{raw \ese data} comprises $2.9k$ source code snippets extracted using \textit{Galeras}\cite{daniel_icsme23} and its library, \textit{SnipGen}\cite{daniel_icse25}. Each snippet is linked to a natural-language description in the \textit{docstring}, along with metadata describing both the snippet and its documentation. The \textit{SnipGen} collects commit-level code snippets from GitHub within a specified time frame. For this case study, we used data that span from January 2, 2022, to January 1, 2023. 

\myp{Procedure}. To formulate a valid \ese causal question, we begin by examining the raw \ese data and characterizing its features (\ie code complexity, keywords, and lines of code) to support the formulation of our question.  The raw data capture four dimensions representing static features derived from the snippets (see \tabref{tab:causal_variables}). The four dimensions are based on the \textit{SnipGen} structure and encode snippet identification, documentation features, syntactic metrics, and maintainability code metrics~\cite{daniel_icse25}. The identification features include only code-related metadata; for instance, \commitID, \filename, and \texttt{path} indicate the snippet location, while \commitMessage and \funName are the \code semantic references. Finally, the snippet consists of \code and \funName. The documentation dimension includes features related to the snippet description, \eg the \docstring is the \code description in natural language; the \docstring has a number of white spaces called \dwhitespaces, and a vocabulary size called \dvocab. Syntactic code metrics comprise descriptive features from code snippets (\eg, \ASTlevels, \identifiers, \ASTnodes). The maintainability code metrics include the lines of code (\nloc) and cyclomatic complexity (\complexity). Although syntactic and maintainability metrics can be extended in future work, they are sufficient for this study and serve as \textit{covariates}. Notice that the covariates have a prefix \texttt{w} , which corresponds to the confounder in the structural causal model explained in \secref{sec:stage1}. \tabref{tab:causal_variables} includes a detailed description for each characteristic.
\begin{table}[h!]
\centering
\resizebox{0.95\linewidth}{!}{%
\begin{tabular}{lclcccl}
\multicolumn{1}{c}{\textbf{Dimension}} &
  \textbf{ID} &
  \multicolumn{1}{c}{\textbf{Feature}} &
  \textbf{Distribution $\mu[\sigma]$} &
  \textbf{Min} &
  \textbf{Max} &
  \multicolumn{1}{c}{\textbf{Description}} \\ \hline
\multirow{7}{*}{\textbf{\begin{tabular}[c]{@{}l@{}}Snippet\\ Identification\end{tabular}}} &
  - &
  \textit{\commitID} &
  - &
   &
   &
  Commit ID from the mined repository \\
 &
  - &
  \textit{repo} &
  - &
   &
   &
  Repository Name refers to the mined project at GitHub \\
 &
  - &
  \textit{path} &
  - &
   &
   &
  The path for the file that includes the \code snippet and the \docstring \\
 &
  - &
  \textit{\filename} &
  - &
   &
   &
  The name of the file that contains the \code snippet \\
 &
  - &
  \textit{\funName} &
  - &
   &
   &
  This is \textit{Function Signature} for the \code snippet, it includes the name and parameters \\
 &
  - &
  \textit{\commitMessage} &
  - &
   &
   &
  The string with the \textit{Commit message} when the file and function was included into the \textit{repo} \\
 &
  - &
  \textit{\code} &
  - &
   &
   &
  The \code block function, it includes the \funName and the function implementation \\ \hline
\multirow{4}{*}{\textbf{Documentation}} &
  - &
  \textit{d\_id} &
  - &
   &
   &
  The assigned \textit{Documentation Id}, this id is linked to the \code snippet \\
 &
  - &
  \textit{\docstring} &
  - &
   &
   &
  The Function documentation as text in natural language \\
 &
  $w_0$ &
  \textit{\dwhitespaces} &
  91.23[143.41] &
  2 &
  1685 &
  The \docstring white spaces count \\
 &
  $w_1$ &
  \textit{\dvocab} &
  32.35[36.42] &
  4 &
  406 &
  The \docstring number words without repetition is formally called vocabulary size \\ \hline
\multirow{7}{*}{\textbf{\begin{tabular}[c]{@{}l@{}}Syntatic\\ Code Metrics\end{tabular}}} &
  $w_2$ &
  \textit{\whitespaces} &
  204.65[366.87] &
  3 &
  14028 &
  Code snippet white spaces \\
 &
  $w_3$ &
  \textit{\wvocab} &
  40.61[35.55] &
  4 &
  421 &
  The \code snippet \textit{vocabulary size}. Vocabulary is the number of words without repetition \\
 &
  $w_4$ &
  \textit{\wnwords} &
  59.68[68.35] &
  4 &
  909 &
  The \code snippet word count with repetition \\
 &
  $w_5$ &
  \textit{\ASTnodes} &
  181.43[187.73] &
  15 &
  3312 &
  The number of nodes from the  \code snippet at the AST representation \\
 &
  $w_6$ &
  \textit{\ASTlevels} &
  12.24[3.09] &
  6 &
  31 &
  The number of AST levels from the \code snippet \\
 &
  $w_7$ &
  \textit{\identifiers} &
  16.55[11.33] &
  1 &
  106 &
  Number of \code identifiers (\eg variable name, function name) \\
 &
  $w_8$ &
  \textit{\ASTerrors} &
  0.09[0.59] &
  0 &
  28 &
  Number of identified AST syntax errors (\eg the \code snippet contains indentation error) \\ \hline
\multirow{3}{*}{\textbf{\begin{tabular}[c]{@{}l@{}}Maintainability \\ Code Metrics\end{tabular}}} &
  $w_9$ &
  \textit{\complexity} &
  3.81[4.36] &
  1 &
  66 &
  The Cyclomatic \code snippet complexity \\
 &
  $w_{10}$ &
  \textit{\nloc} &
  18.06[20.60] &
  1 &
  548 &
  The \code snippet number of lines of code \\
 &
  $w_{11}$ &
  \textit{\tokenCount} &
  112.21[122.27] &
  6 &
  2133 &
  Number of tokens in the \code snippet
\\ \bottomrule
\end{tabular}%
}

\caption{Identified raw dataset properties classified in four dimensions with their description}
\label{tab:causal_variables}
\end{table}

After identifying the dataset's features, we conduct an exploratory analysis to examine their frequencies and distributions. \tabref{tab:causal_variables} depicts the $\mu$-mean, $\sigma$-standard deviation, minimum and maximum values for each feature. This descriptive analysis helps us to later design the correlation analysis  (see \ref{sec:stage1}) and classify the characteristic as the corresponding causal variable (\ie instrumental variable, effect modifier, or confounder) as shown in \figref{fig:causalgraph}.

Our formulated \textit{causal question} is to \emph{to what extent do prompt engineering features (\ie length, \wvocab, \complexity, etc.) influence the code completion performance of an LLM?} The question is especially relevant given that recent studies have demonstrated performance gains from prompt engineering without retraining \llm~\cite{ji_benchmarking_2023}. Exploratory data analysis helps us to identify whether the data set supports our causal question. In this case, we identify that we can formulate prompt variations by intervening in the \code and merging the \docstring and \code to construct the prompts. In addition, \code and \docstring features can be used as confounders.

\myp{Outputs.} As output from this stage, an exploratory data analysis of our raw data \ese and a refined causal question supported by the \ese dataset.

\subsection {[Stage$_1$] Modeling Causal SE Problem} \label{sec:stage1}
\myp{Inputs.}  From our previous Stage$_0$, we obtained the data features and distributions to answer the formulated causal question. To implement this case study, we used \textit{DoWhy}, a Python library that helps identify various causal estimands given the specified \scm. We also used our knowledge on \ese to validate the proposed correlations of variables and classifications of causal variables.

\myp{Procedure.} To model the causal SE problem using \scm, we explore multiple \textit{DAG} candidates that align with our data and the formulated causal question. The \textit{causal discovery} process helps us to evolve an initial DAG. The purpose of causal discovery is to define a stable \scm given the identified features and their relation to answer the causal questions. The evolution DAG is depicted in \figref{fig:causalgraph}. Each DAG is tested to select the final \scm. 

The initial DAG model $m_1$ is based on the basic causal graph triangle depicted in \figref{fig:causalgraph}. The features listed in stage$_0$ (see \secref{sec:stage0}) are categorized as common causes $Z$ because they affect both the input prompt $T$ and the result $Y$. The process requires some assumptions about the data, informed by both domain expertise and results from the previous exploratory analysis. To confirm our assumptions, we need to run experiments by configuring the treatments -- with the prompt variations-- and collecting the model outcomes.

The causal question concerns the prompt's impact on code generation. Thus, each treatment $T$ is an intervention of the input prompt. The proposed case study consists of three treatments to assess the effects of the prompt and establish the experiments (see \tabref{tab:treatments}). The first \textbf{Treatment $T_0$} is the baseline and comprises the instruction and the \randomCut sequence. The \randomCut randomly removes a line after the signature of the snippet, ensuring a meaningful starting code and context to complete the code. The second \textbf{Treatment $T_1$} includes only the snippet signature; this treatment reduces the information content while assessing the importance of the signature. Treatment $T_1$ restricts the output by asking one to remove any comments or descriptions about the generated code. Finally, the third \textbf{Treatment $T_2$} extends the instruction description and comprises three snippet features (\ie signature, random cut code, and docstring). Treatment $T_2$ offers comprehensive information regarding intentionality and expected output while imposing the same restrictions as $T_1$ and requesting only code as the result.

\begin{table*}[!h]
\centering

\resizebox{\linewidth}{!}{%
\setlength{\tabcolsep}{5pt} 
\begin{tabular}{clp{7.5in}}

\multicolumn{1}{c}{\textbf{ID}} &  & \multicolumn{1}{c}{\textbf{Prompt Template}} \\ \hline
$T_0$ &  & Complete the following \textless{}\textit{language}\textgreater method: \textless{}\randomCut\textgreater{} \\
$T_1$ &  & Write a \textless{}\textit{language}\textgreater method that starts with \textless{}\textit{signature}\textgreater{}, I need to complete this function. Remove comments, summary, and descriptions \\
$T_2$ &  & Remember you have a \textless{}\textit{language}\textgreater function names  \textless{}\textit{signature}\textgreater, the function starts with the following code  \textless{}\randomCut\textgreater. The description for the function is: \textless{}\docstring\textgreater{}; remove comments; remove summary; remove description; Return only the code.\\ 

  \bottomrule
\end{tabular}
\vspace{-0.3cm}
}
\caption{Prompt templates for code completion.}
\label{tab:treatments}

\end{table*}

After designing the treatments, we collect the results of the model $Y$. To gather $Y$ outcomes, we feed each set of treatments $T$ as input for the OpenAI GPT-3 model~\cite{brown2020languagemodelsfewshotlearners} and then capture the outcome. After collecting $Y$ outcomes, we computed the associated features for each treatment \ie the \inwords, \iwhitespaces, and \ivocab for the input prompts for the treatments $T_0,T_1$, and $T_2$, and outcome \ie \ewhitespaces, \enwords, and \evocab. \tabref{tab:i-o-metrics} depicts the metrics for each $T$ and $Y$. We observe larger word counts for $T_2$ than for $T_0$ and $T_1$. Vocabulary size for $T_0$ and $T_1$ are similar, while $T_2$ duplicates the size. The input prompt metrics have the prefix $i$ indicating the assigned categorization and an \textit{instrumental variable}, while the outcomes $Y$ feature metrics have the $e$ prefix indicating an \textit{effect modifier}.
\begin{table}[h!]
\centering
\resizebox{\linewidth}{!}{%
\begin{tabular}{lllllrrrlrrrlrrr}
\multicolumn{1}{c}{} &
   &
  \multicolumn{1}{c}{} &
   &
   &
  \multicolumn{3}{c}{\textbf{$T_0$ = Control}} &
  \multicolumn{1}{c}{\textbf{}} &
  \multicolumn{3}{c}{\textbf{$T_1$}} &
  \multicolumn{1}{c}{\textbf{}} &
  \multicolumn{3}{c}{\textbf{$T_2$}} \\ \cline{6-8} \cline{10-12} \cline{14-16} 
\multicolumn{1}{c}{\textbf{}} &
  \textbf{ID} &
  \multicolumn{1}{c}{\textbf{Features}} &
  \multicolumn{1}{c}{\textbf{Description}} &
  \multicolumn{1}{c}{} &
  \multicolumn{1}{c}{\textbf{$\mu[\sigma]$}} &
  \multicolumn{1}{c}{\textbf{min}} &
  \multicolumn{1}{c}{\textbf{max}} &
  \multicolumn{1}{c}{\textbf{}} &
  \multicolumn{1}{c}{\textbf{$\mu[\sigma]$}} &
  \multicolumn{1}{c}{\textbf{min}} &
  \multicolumn{1}{c}{\textbf{max}} &
  \multicolumn{1}{c}{\textbf{}} &
  \multicolumn{1}{c}{\textbf{$\mu[\sigma]$}} &
  \multicolumn{1}{c}{\textbf{min}} &
  \multicolumn{1}{c}{\textbf{max}} \\ \cline{1-4} \cline{6-8} \cline{10-12} \cline{14-16} 
\multirow{3}{*}{\textit{\textbf{\begin{tabular}[c]{@{}l@{}}Input Prompt\\ Feature Metrics\end{tabular}}}} &
  $i_0$ &
  \textit{\iwhitespaces} &
  Number of spaces in the prompt &
   &
  114.52[217.59] &
  6 &
  8528 &
   &
  128.52[217.59] &
  20 &
  8542 &
   &
  225.13[269.66] &
  40 &
  8650 \\
 &
  $i_1$ &
  \textit{\inwords} &
  Number of words with repetition &
   &
  38.90[42.69] &
  7 &
  622 &
   &
  52.90[42.69] &
  21 &
  636 &
   &
  110.51[86.97] &
  38 &
  1079 \\
 &
  $i_2$ &
  \textit{\ivocab} &
  Number of words without repetition &
   &
  30.33[24.05] &
  7 &
  241 &
   &
  43.94[23.57] &
  21 &
  252 &
   &
  76.26[43.93] &
  30 &
  457 \\ \cline{1-4} \cline{6-8} \cline{10-12} \cline{14-16} 
\multirow{3}{*}{\textit{\textbf{\begin{tabular}[c]{@{}l@{}}Generated Outcome\\ Feature metrics\end{tabular}}}} &
  $e_0$ &
  \textit{\ewhitespaces} &
  Number of generated of whitespaces &
   &
  160.11[204.31] &
  0 &
  7257 &
   &
  207.77[231.69] &
  0 &
  7204 &
   &
  121.61[167.60] &
  5 &
  6554 \\
 &
  $e_1$ &
  \textit{\enwords} &
  Number of generated words &
   &
  84.74[69.70] &
  1 &
  652 &
   &
  113.32[72.98] &
  1 &
  767 &
   &
  60.46[40.11] &
  4 &
  480 \\
 &
  $e_2$ &
  \textit{\evocab} &
  Number of words without repetition &
   &
  56.18[38.67] &
  1 &
  298 &
   &
  72.50[36.90] &
  1 &
  294 &
   &
  45.90[24.17] &
  4 &
  197 \\ \hline
\end{tabular}%
}

\caption{Feature metrics for the input prompts $T$ and generated outcomes $Y$}
\label{tab:i-o-metrics}
\end{table}

\begin{figure*}[h]
\centering
\begin{subfigure}[t]{0.49\textwidth}
    \centering
    \includegraphics[width=\linewidth]{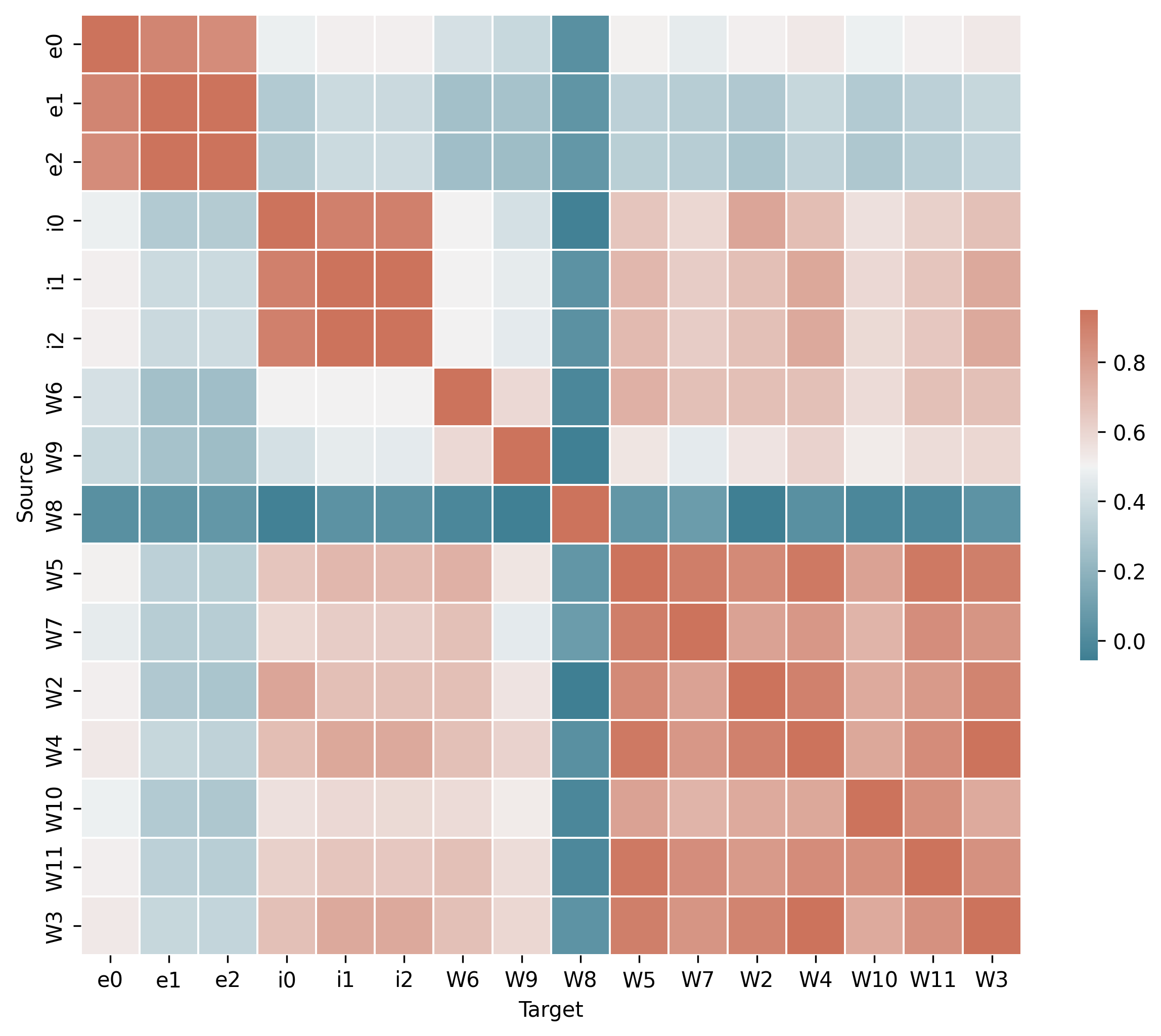}
    
    \label{fig:control-spearman}
\end{subfigure}
\hfill
\begin{subfigure}[t]{0.49\textwidth}
    \centering
    \includegraphics[width=\linewidth]{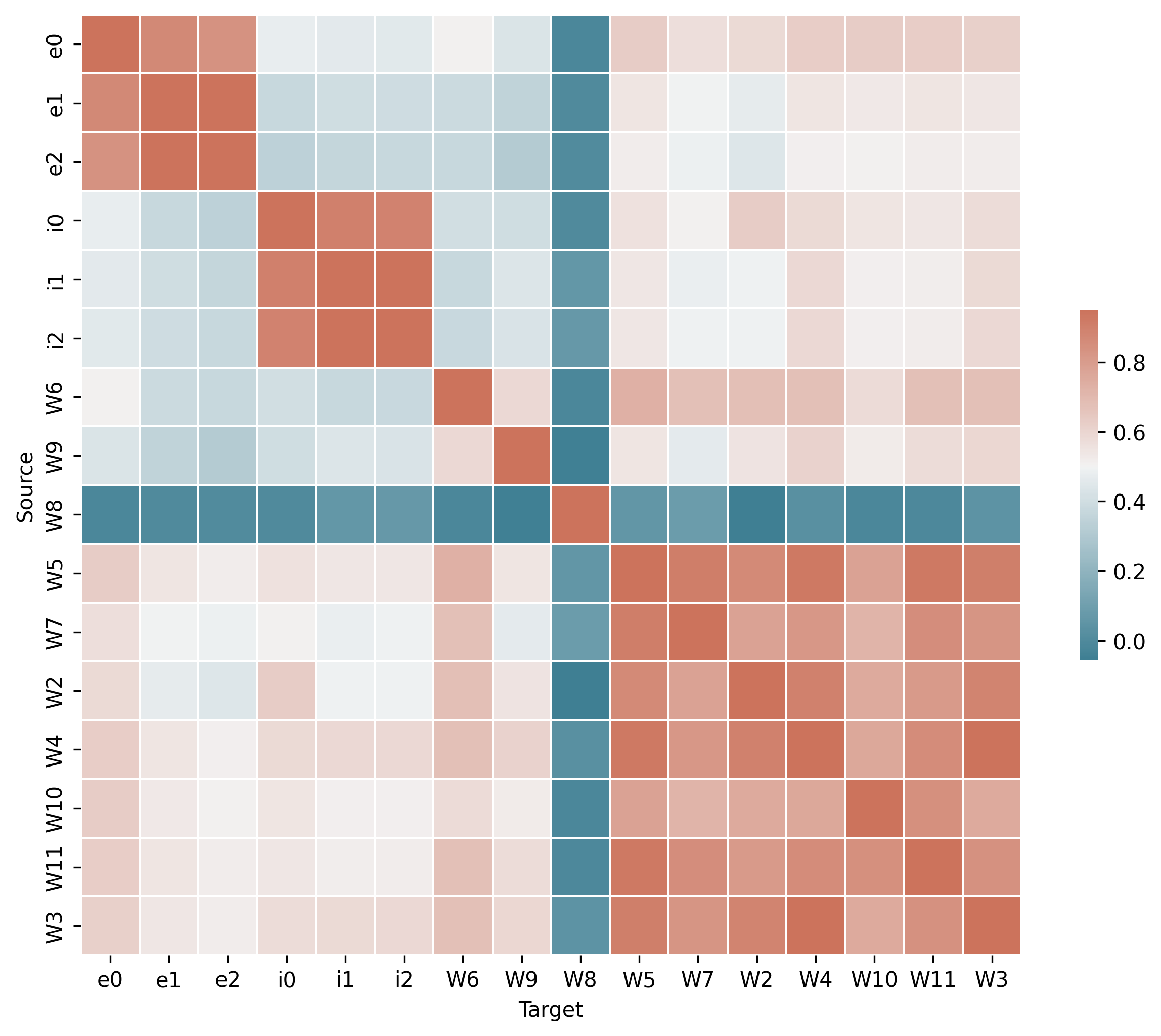}
    
    \label{fig:sub2}
\end{subfigure}
\caption{Spearman $\rho$ correlation for instrument variables ($i$), confounders ($w$), and effect modifiers $(e)$. \textbf{Left:} $T_0$ control covariates correlation. \textbf{Right:} $T_2$ covariates correlation.}
\label{fig:correlations}
\end{figure*}

The execution of treatments and the capture of the resulting outcomes are part of the \textit{data collection} process described in the \causalse framework (see \figref{fig:causalSEpipeline}). The newly collected data contains $8.77k$ data points. Each data point is then curated and restructured in the \textit{data transformation}. Data transformation consists of creating a \textit{structured data} set that includes the identified confounders $Z=w_n$, instrumental variables $i$, and effect modifiers $e$ for each treatment $T$ and the result generated $Y$. Using the identified causal variables $w,i$, and $e$ we create Pearson $r$ and Spearman $\rho$ correlation. \figref{fig:correlations} depicts confusion matrices using $\rho$ and covariates. Here, the correlation is performed using the frequencies from \tabref{tab:causal_variables} and \tabref{tab:i-o-metrics}.

\begin{wrapfigure}{r}{0.5\linewidth}
 \centering
 \includegraphics[width=\linewidth]{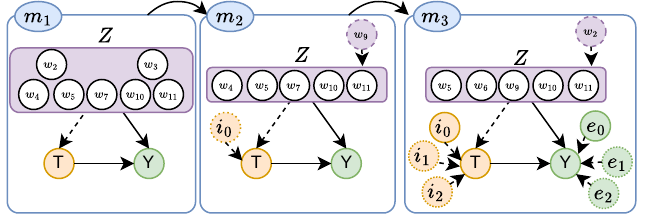}
 \caption{Structural Causal Model Evolution from $m_1$ to $m_3$.}
 \label{fig:causalVariables}
\end{wrapfigure}

We observe that \ASTnodes, \identifiers, \whitespaces, \wnwords, \nloc, \tokenCount, and \wvocab are highly correlated, which confirms our assumptions for the first iteration. To complete the first iteration, the generated DAG needs to be estimated and \textit{vetted}. The estimation and vetting require the subsequent stages to complete the \causalse cycle. Up to now, the first iteration generated the first \scm candidate, $m_1$ (see \figref{fig:causalVariables}).

After collecting more variables and observing their relationships, \causalse enables the exploration of additional DAG models and updates the current model to obtain the final \scm. Therefore, given the correlations and new variables, we update $m_1$ by updating and filtering the $Z$ confounders (\ie \whitespaces, \wvocab, \wnwords, \ASTnodes, \identifiers, and \nloc), and propose $m_2$ and $m_3$. $m_2$ is a model that includes instrumental variables (\eg \ivocab, \inwords) and $m_3$ include both instrumental variables and effect modifiers (\eg \evocab, \enwords). As a result, we obtain three DAG candidates for testing and vetting to obtain a \scm.

\myp{Outputs.} The outputs from this stage comprise (i) the structured dataset that contains the treatment and the model outcomes for each treatment, and (ii) a set of \scm candidates with DAG containing proposed variables (\ie \textit{confounders,} \textit{instrumental variables}, \textit{effect modifiers}).

\myp{Discussion.} \figref{fig:correlations} (left) illustrates confounders $w$ with low impact on effect modifiers $e$ within the control treatment. In contrast, the analysis for treatment $T_2$ reveals a slightly higher impact of $w$ on both the $e$ and $i$ variables. We select confounders for the Structural Causal Model (\scm) based on these observed correlations; for example, while \wnwords demonstrates a correlation with $e$ and $i$, variables such as \ASTlevels ($w_6$), \ASTerrors ($w_8$), and \complexity($w_9$) show negligible impact and are therefore excluded from the initial candidate DAG. However, \scm will undergo further validation in subsequent stages. If iterations of \causalse --following effect estimation and graph vetting-- suggest a missing confounder, we can incorporate additional variables or extra new features. In particular, practitioners may still find value in monitoring the impact of \complexity during the code generation phase, regardless of its initial exclusion.

\subsection{[Stage$_2$] Identifying causal Estimand:}
\myp{Inputs.} This stage aims to derive the mathematical expression that represents the causal effect. This stage takes as input first, the candidate \scm $m_1,m_2, m_3$ defined in Stage$_1$, including their treatment $T$, outcome $Y$. Second, the binary treatment structure, where $T_0$ is the \textit{control} prompt template and $T_1$ and $T_2$ are alternative prompt strategies. Finally, the structured dataset with confounders $Z=w$, the instrumental variables $i$, and effect modifiers $e$ computed for each prompt-completion pair.

\myp{Procedure. } We formalize the causal question as an stimand dedicing whether the effect of prompt strategy on code quality should be expressed as the Average Treatment Effect (ATE) or Conditional Average Treatment Effect (CATE) over specific covariates. For each \scm, we use DoWhy's identification module to determine the appropriate adjustment strategy (\ie backdoor, frontdoor, or instrumental variable adjustments) given the graph structure. Because the experiments are binary, we encode comparisons as $T_0$ vs. $T_1$ and $T_0$ vs. $T_2$, and derive the corresponding symbolic expression for $p(Y\mid do(T=t))$ using the identified adjustment set. To reduce stochastic variance in \llm outputs, we generate $5$ completions per prompt and average the evaluation metrics before plugging them into the estimand. This reduces random variance in Levenshtein distance estimates and ensures that the estimated effects reflect systematic differences in prompts rather than sampling noise.

\myp{Outputs. } The output from this stage is (i) a specification of whether backdoor adjustment or instrumental variable identification is used for each model configuration and (ii) a pre-processed dataset of averaged results ready for numerical estimation in Stage$_3$.

\subsection{[Stage$_3$] Causal effect estimation:} \label{sec:case_s3}
This stage quantifies the causal effect of the treatment on the outcome using the identified estimand from stage$_2$. 
\myp{Inputs. } This stage consumes first the identified estimands for each \scm and treatment contrast form Stage$_2$, second the structured dataset with treatment indicators, averaged outcome metrics (\eg Levenshtein distance), and covariates to be used in adjustment, and finally the choice of estimation methods. To control for confounding variables and estimate the causal effect of prompt treatments on model performance metrics, we use statistical estimation methods - \textit{propensity score matching (PSM)}, \textit{propensity score stratification (PSS)}, and \textit{propensity score weighting (PSW)}. 

\myp{Procedure. } For \scm $m_1$, we estimate the ATE using propensity-score-based backdoor adjustment. We fit a propensity model for treatment assignment, then apply PSM, PSS, and PSW to obtain effect estimates for $T_0$ vs. $T_1$ and $T_0$ vs. $T_2$. For \scm $m_2$ and $m_3$, which include instrumental and effect-modifier variables, we employ DoWhy's estimators to quantify the causal effect using the specified instruments. For each estimator and comparison, we compute the mean causal effect and p-values, interpreting positive values as an increase in Levinshtein distance (worse predictions) and negative values as a reduction in distance (better predictions) relative to the control.
\begin{table*}[!ht]

\centering

\resizebox{\linewidth}{!}{%
\begin{tabular}{llllrrrrlrrrr}
\multicolumn{3}{c}{\textbf{Causal Analysis Setup}} &  & \multicolumn{2}{r}{\textbf{Estimation(T$_0$, T$_1$)}} & \multicolumn{2}{r}{\textbf{Validation(T$_0$, T$_1$)}} &  & \multicolumn{2}{r}{\textbf{Estimation(T$_0$, T$_2$)}} & \multicolumn{2}{r}{\textbf{Validation(T$_0$, T$_2$)}} \\
\multicolumn{1}{c}{\textit{\textbf{SCM}}} & \multicolumn{1}{c}{\textit{\textbf{Method}}} & \multicolumn{1}{c}{\textit{\textbf{Estimand}}} & \multicolumn{1}{c}{} & \multicolumn{1}{c}{\textit{\textbf{Mean}}} & \multicolumn{1}{c}{\textit{\textbf{p-value}}} & \multicolumn{1}{c}{\textit{\textbf{Placebo}}} & \multicolumn{1}{c}{\textit{\textbf{RCR}}} & \multicolumn{1}{c}{} & \multicolumn{1}{c}{\textit{\textbf{Mean}}} & \multicolumn{1}{c}{\textit{\textbf{p-value}}} & \multicolumn{1}{c}{\textit{\textbf{Placebo}}} & \multicolumn{1}{c}{\textit{\textbf{RCR}}} \\ \cline{1-3} \cline{5-8} \cline{10-13} 
\multirow{3}{*}{$m_1$} & PSM & backdoor &  & 4.2E-03 & 4.2E-01 & -1.3E-04 & 9.0E-03 &  & -9.8E-03 & 2.9E-01 & -6.9E-05 & -8.2E-03 \\
 & PSS & backdoor &  & 9.0E-03 & 1.0E-03 & 8.8E-05 & 9.1E-03 &  & -8.3E-03 & 1.0E-03 & 6.8E-05 & -8.3E-03 \\
 & PSW & backdoor &  & 9.0E-03 & 1.0E-03 & -5.9E-05 & 9.0E-03 &  & -8.3E-03 & 1.0E-03 & -7.4E-05 & -8.3E-03 \\
$m_2$ & IV & iv &  & 1.4E-01 & 1.0E-03 & -3.0E-04 & 1.4E-01 &  & 1.4E-01 & 1.0E-03 & 3.5E-02 & 3.6E-02 \\
$m_3$ & IV & iv &  & 1.4E-01 & 1.0E-03 & 5.9E-04 & 1.4E-01 &  & 1.4E-01 & 1.0E-03 & 3.6E-02 & 3.6E-02 \\
\bottomrule
\end{tabular}%
}
\caption{Causal effect estimation and validation with placebo and Random Cause Refutation (RCR). IV = Instrumental Variable method in DoWhy. $\mu(T_0)=0.038$, $\mu(T_1)=0.0$}
\label{tab:causal-validation}
\vspace{-0.4cm}
\end{table*}

\myp{Outputs. } The stage first reports the numerical ATE estimates and associated p-values for each \scm, estimator, and treatment comparison, summarized in \tabref{tab:causal-validation}. Second, characterization of the direction and magnitude of the effect (null, positive, or negative) for each prompt treatment. Finally, the input effect estimates are to be stress-tested in Stage$_4$ through refutation analyzes.

\myp{Discussion. } \tabref{tab:causal-validation} shows the causal effect estimation comparing $T_0$ control against $T_1$ and against $T_2$. \tabref{tab:causal-validation} also presents the mean and p-values for each of the \scm. $m_1$ with PSM makes a backdoor estimation with a mean of $0.0042$ and p-value $0.42$, which indicates \textit{no statistically significant evidence} with a low treatment effect. However, we observe a low impact between $T_0$ and $T_2$ with a PSM mean of $-0.0098$ and a p-value of $0.286$. The PSS and PSW on $m1$ indicate a mean of $0.009$ versus $T_1$ and $-0.0082$ with a p-value of $0.001$ for both treatments. These PSS and PSW suggest a slight impact from treatment. For models $m_2$ and $m_3$, we observe a mean of $\approx 0.13$ in both treatments and models. The effect on the computation is inconclusive for these two models and no variability was observed between treatments. %

\subsection{[Stage$_4$] Causal effect validation:} Refutation methods help validate the robustness of causal estimates by testing how sensitive the results are to different interventions in the data or assumptions \scm.

\myp{Inputs. } This stage takes as input the estimated causal effect (ATEs), and the corresponding \scm configuration from the Stage$_3$. The original structured dataset and \scm specifications required to generate refuted variants, and the configuration of refuters provided by DoWhy, a Python library for running \docode, provides two refuters. In this case study, we apply the \textit{placebo treatment refuter} and the \textit{random common cause refuter (RCR)}. 

\myp{Procedure. } We first apply the placebo treatment refuter, replacing the true treatment indicator with a randomly permuted ``placebo'' variable and re-estimating the effect under the same estimator and \scm; we expect estimates close to zero if the original pipeline is not overfitting noise. Next, we apply the RCR, injecting a synthetic random covariate as an additional confounder $Z$ and recomputing the effect to assess sensitivity to unobserved confounding. For each configuration (\scm, estimator, contrast of treatment), we compare the placebo and random-cause estimates against the original mean effect to judge robustness.

\myp{Outputs. } The outputs include the placebo and RCR effect estimates for all \scm-stimator combinations reported alongside the original means in \tabref{tab:causal-validation}. A qualitative assessment of PSS and PSW against PSM. A validated set of causal estimates deemed sufficiently robust to inform the explanatory analysis in Stage$_5$.

\myp{Discussion. } We observe almost zero placebo effect on all validations, which means that the estimation is valid. However, some variations in the RCR compared to the estimated mean, for example $0.009$ for $m_1$ with a mean of $0.004$ between $T_0$ and $T_1$. Therefore, we consider the more robust validation methods for PSS and PSW in this case, since the difference is zero when comparing $T_0$ and $T_1$, and $T_0$ and $T_2$. Structural Causal Models $m_2$ and $m_3$ have the expected result on refuters, indicating that more uncaptured and hidden variables are required for a more precise \scm.

\subsection{[Stage$_5$] Producing Explanations:}
\myp{Inputs. }This stage builds on the validated causal effect estimates and refutation results from Stage$_3$ and Stage$_4$. As an input, this stage requires the ATE and CATE estimates, the \scm, and the structured dataset. %

\myp{Procedure. } In this stage, we first conduct an \textit{associational analysis} by comparing metric distributions for $T_0$, $T_1$, and $T_2$ and computing rank correlations and divergence measures to characterize the observed differences between the prompt strategies. Next, we contrast these descriptive patterns with the interventional results. Then we interpret the ATE and CATE estimates (and their signs) under each SCM and estimator to answer the question, “What if we change the prompt strategy?” Finally, we synthesize a narrative explanation that highlights the discrepancies between association and causation, emphasizing when correlational improvements fail to translate into statistically supported causal effects. The explanation aims to describe the impacts of prompt engineering in practical settings.

\textbf{Associational Analysis.} 
The associational analysis focuses on how each treatment's metric measures against the ground truth. \tabref{tab:bootstrap-results} depicts the bootstrapped mean for the normalized Levenshtein distance, cosine similarity, and CodeBLUE metrics in relation to control $T_0$, and treatments $T_1$ and $T_2$. Using the normalized Levenshtein distance metric, the associational analysis shows a positive impact of $T_2$ relative to the control treatment and a negative impact of $T_1$. The cosine similarity and CodeBLEU are unclear; the medians are similar, and the scaled ranges for each treatment show no difference. To analyze this correlation, we also computed the Spearman correlation and Jensen-Shannon divergence (JSD) between the treatments. We observe Spearman correlations of 0.402 between $T_0$ and $T_1$, and 0.464 between $T_0$ and $T_2$. The Spearman results suggest a moderate positive monotonic relationship between $T_0$ and $T_1$, with a slightly stronger relationship with $T_2$. The observed JSD for $T_0$ and $T_1$ of 0.112, $T_0$ and $T_2$ of $0.185$ and 0.286 for $T_1$ and $T_2$. The last value indicates the highest divergence, suggesting that these two treatments lead to greater variability. 

\definecolor{color1bg}{HTML}{9698ED}
\begin{table}[!h]
   
\centering
\scalebox{0.98}{%
\begin{tabular}{llcclcclcc}
\multicolumn{1}{c}{} &
   &
  \multicolumn{2}{c}{\textbf{Normalized Lev. Distance}} &
   &
  \multicolumn{2}{c}{\textbf{CodeBleu}} &
   &
  \multicolumn{2}{c}{\textbf{Cosine Similarity}} \\
\multicolumn{1}{c}{\textit{\textbf{Treatment}}} &
   &
  \textit{\textbf{$\mu$}} &
  \textit{\textbf{$\sigma$}} &
   &
  \textit{\textbf{$\mu$}} &
  \textit{\textbf{$\sigma$}} &
   &
  \textit{\textbf{$\mu$}} &
  \textit{\textbf{$\sigma$}} \\ \cline{1-1} \cline{3-4} \cline{6-7} \cline{9-10} 
\textit{control} &  & 4.81E-02 & 8.60E-04 &  & 4.78E-01 & 4.06E-03 &  & 5.26E-01 & 4.40E-03 \\
\textit{T1}      &  & 5.71E-02 & 8.90E-04 &  & 4.57E-01 & 4.00E-03 &  & 4.34E-01 & 4.63E-03 \\
\textit{T2} &
   &
  \cellcolor[HTML]{9698ED}{\color[HTML]{FFFFFF} 3.98E-02} &
  8.70E-04 &
   &
  \cellcolor[HTML]{9698ED}{\color[HTML]{FFFFFF} 4.86E-01} &
  4.09E-03 &
   &
  \cellcolor[HTML]{9698ED}{\color[HTML]{FFFFFF} 5.77E-01} &
  4.65E-03 \\ \bottomrule
\end{tabular}%

}
\caption{Bootstrapped (n=500) mean ($\mu$) and STD ($\sigma$) for normalized Levenshtein distance, CodeBLEU, and Cosine Similarity for each treatment outcome $Y$. \textcolor{color1bg}{Highligthed} indicates the best score.}

\label{tab:bootstrap-results}
\end{table}

\textbf{Interventional Analysis.} 
The interventional analysis results presented in \secref{sec:case_s3} and \tabref{tab:causal-validation} reveal that, for model $m_1$, the propensity score analyses do not consistently align with our initial hypotheses. Although \textit{PSM} frequently yielded statistically insignificant effects, the \textit{PSS} and \textit{PSW} methods consistently suggested impacts that partially support our expectations. Specifically, $T_1$ resulted in a marginal increase in distance, while $T_2$ led to a slight reduction. Ultimately, these findings indicate that $T_1$ and $T_2$ did not perform as expected relative to the control for model $m_1$, as the additional context of $T_2$ did not produce a more significant causal effect.

\textbf{Comparative Analysis.}

While the interventional analysis using the propensity score methods suggested that prompt treatments $T_1$ and $T_2$ had only a slight, unanticipated causal impact on model $m_1$'s distance metrics, the associational analysis presented no meaningful statistical differences for cosine similarity and CodeBLEU. This contrast highlights the distinction between observed relationships and actual causal effects, indicating that, while some associations exist, the direct influence of treatments was less significant than hypothesized. When associational results suggest improvements but interventional estimates and refuters do not, practitioners using \causalse need to revise the \scm or data collection and return to Stage$_0$.

\myp{Outputs. } The output of this stage comprises the \textit{associational analysis}, the \textit{interventional conclusions}, and a \textit{comparative analysis} that positions \causalse as a guardrail against over-interpreting correlational improvements, recommending null-effect conclusions and \scm refinement when interventional and associational results diverge.

\myp{Discussion:} This case study evaluates the influence of prompt engineering variations on model accuracy. Our results indicate a marginal but statistically significant improvement—confirmed via bootstrapping—when utilizing $T_2$ with the maximum context window, as measured by Levenshtein distance. While raw distance scores remain low due to normalization, the statistical distinction persists. To validate these interventional correlations, we applied causal falsification methods, which revealed only a minimal causal impact. Furthermore, canonical metrics such as CodeBLEU and Cosine Similarity yielded inconclusive results, suggesting that observed performance gains are highly sensitive to the choice of evaluation metric

\subsection{Threats to Validity}

\textbf{Model randomness.} LLM outputs are stochastic. While we reduced variance by sampling multiple generations, future work should explore causal estimators that explicitly model outcome uncertainty (\eg bootstrap CATE intervals).

\textbf{Prompt–model coupling.} Prompt effectiveness may vary across models (\eg GPT-3 vs GPT-4). Our findings are demonstrated in GPT-3; generalization requires replication in multiple LLMs. \causalse itself is model-agnostic, but we note this as an external validity threat.

\textbf{Training data overlap.} GPT-3 training data reportedly cover sources up to 2021. Our dataset spans Jan 2022–Jan 2023, which reduces—but does not eliminate—the risk of training data overlap. Potential leakage could inflate causal estimates if prompts resemble training data. We flag this as an unavoidable threat to validity in LLM-based ESE studies.

\textbf{Unobserved or missing confounders.} \causalse assume that relevant confounders can be identified and measured. If important confounders are tacit or missing, treatment effect estimates may be biased. We mitigate this risk through (i) expert-guided graph design, (ii) sensitivity checks using refuters (\eg placebo, random common cause), and (iii) iterative refinement of the study design and data collection. Nevertheless, as in all causal inference, the possibility of unobserved confounding remains a limitation and should be explicitly reported in empirical SE studies.

\section{Related Work}\label{sec:related}

Empirical software engineering (\ese) has traditionally relied on correlation-based statistical methods, often neglecting causal inference. However, there is growing recognition that causal methods -such as directed acyclic graphs (\dags) and $do$-calculus- can uncover true causal relationships and improve the transparency and interpretability of research findings~\cite{neto_evolution_2019}. Despite this potential, causal inference remains underused in SE. Surveys show that few studies report that fewer than 2\% of articles reported controlled experiments with deliberate interventions, and that the reporting was often vague, unsystematic, and hampered by inconsistent terminology, controlled interventions, or adopting clear causal reasoning~\cite{sjoeberg_survey_2005}.

Causal methods are especially valuable in areas where controlled experiments are complex, including software maintenance, quality assurance, and performance analysis~\cite{chadbourne_applications_2023, siebert_applications_2023}. Case studies demonstrate their effectiveness in tasks such as vulnerability detection~\cite{rahman_towards_2024}, fault localization~\cite{kucuk_improving_2021}, and microservice failure tracking~\cite{pham_root_2024}. Causal models have also provided deeper insight than correlational ones in studies of programming languages~\cite{furia_towards_2024, furia_mitigating_2025} and architectural flaws~\cite{kazman_causal_2017}. Recent advances extend causal inference to the generation of code based on \llm. Frameworks like CodeSCM~\cite{gupta_codescm_2025} and Ji et al.’s benchmark~\cite{ji_benchmarking_2023} use structural causal models to evaluate the influence of prompts and detect memorization, leading to improved prompt design and model transparency.

Although adoption is still limited, causal inference offers a principled path to understanding \textit{why} software systems behave as they do, supporting better decisions throughout the development lifecycle and aligning with an ongoing call for standardized, reproducible causal analysis in \ese.

\section{\causalse Discussion and Future Research}
\label{sec:discussion}

The integration of causal philosophical perspectives into empirical software engineering (ESE) research provides a foundational framework for addressing the question: \textit{How do software researchers enable causal inference in their empirical evaluations?} Our descriptive case study on the impact of prompt engineering on the generation of LLM codes illustrates how adopting Pearl’s interventionist causality, rooted in structural causal models (\scms), resolves the limitations of traditional statistical methods, \ie \textit{confounding bias}, while aligning with the mechanistic philosophical views of causation. 

Enabling causal inference in \ese requires formalizing assumptions through \scm and applying do-calculus to simulate interventions. Our study demonstrates this by contrasting associational outcomes (\ie amount of prompt information appearing related to better code prediction) with interventional results using propensity score matching to control for confounders such as input complexity. By explicitly defining \da and applying the operator \eg $P(CodePrediction|do(PromptStrategy))$, we isolate the true effect of treatments (prompt strategies) from confounding variables. This aligns with Pearl's causation ladder, in which answering what if questions (L$_2$) requires explicit causal assumptions absent in purely statistical models. It should be clarified that comprehensive empirical experimentation with extended groups of \llms and testbeds was beyond the scope of our case study. %

\textbf{How can software engineering problems be solved with \causalse?} The formulation of the hypothesis refines the research objectives and clarifies the experimental goals~\cite{felderer_evolution_2020}, requiring appropriate testing methods based on the research questions. Empirical software evaluation began with frequentist statistical analysis, where Arcuri \etal provided guidelines for handling variability in randomized algorithms~\cite{Arcuri_2011, Arcuri_2014}. Furia et al. later advanced Bayesian data analysis as a more flexible and interpretable alternative~\cite{Furia_2019, torkar2020bayesiandataanalysisempirical}, offering direct probability statements about parameters. As \llms influences SE research, the hypotheses have changed from basic performance comparisons to complex questions involving vague instructions or hypothetical scenarios. This paper introduces causal analysis as a rigorous framework to address interventional and counterfactual questions in software engineering.

\textbf{What Software Engineering problems can be solved with \causalse?} We have identified some common software engineering problems that can be solved with \causalse(\eg code generation task). Although we present a comprehensive classification of research questions grounded in authors' experiences and a preliminary literature review, we invite practitioners and software engineering researchers to expand it based on their empirical needs. \causalse is not limited to LLM evaluation; it can be extended to support hypotheses that require interventions in a wider range of ESE contexts. Moreover, by incorporating counterfactual analysis, \causalse enables the exploration of hypothetical and retrospective questions.

\bibliographystyle{ACM-Reference-Format}
\bibliography{util/bibliography}

\appendix

\end{document}